# MESOSCALE SOLUBILIZATION AND CRITICAL PHENOMENA IN BINARY AND QUASI-BINARY SOLUTIONS OF HYDROTROPES

Andreas E. Robertson[a,b], Dung H. Phan[b], Joseph E. Macaluso[b], Vladimir N. Kuryakov[c], Elena V. Jouravleva[d], Christopher E. Bertrand[e], Igor K. Yudin[c], Mikhail A. Anisimov[b,d,*]

[a]*Montgomery Blair High School, 51 University Boulevard East, Silver Spring, MD 20901, USA*

[b]*Department of Chemical and Biomolecular Engineering, University of Maryland, College Park, MD 20742, USA*

[c]*Oil and Gas Research Institute of the Russian Academy of Sciences, Gubkina 3, Moscow, 117333 Russia*

[d]*Light Scattering Center, Institute for Physical Science & Technology, University of Maryland, College Park, MD 20742, USA*

[e]*NIST Center for Neutron Research, National Institute of Standards and Technology, 100 Bureau Drive, Gaithersburg, MD 20899, USA*

*Corresponding author at: Department of Chemical & Biomolecular Engineering, University of Maryland, College Park, MD 20742, USA. *Email address:* anisimov@umd.edu.

**Abstract**

Hydrotropes are substances consisting of amphiphilic molecules that are too small to self-assemble in equilibrium structures in aqueous solutions, but can form dynamic molecular clusters H-bonded with water molecules. Some hydrotropes, such as low-molecular-weight alcohols and amines, can solubilize hydrophobic compounds in aqueous solutions at a mesoscopic scale (~100 nm) with formation of long-lived mesoscale droplets. In this work, we report on the studies of near-critical and phase behavior of binary (2,6-lutidine–$H_2O$) and quasibinary (2,6-lutidine–$H_2O$/$D_2O$ and tert-butanol/2-butanol–$H_2O$) solutions in the presence of a solubilized hydrophobic impurity (cyclohexane). In additional to visual observation of fluid-phase equilibria, two experimental techniques were used: light scattering and small-angle neutron scattering. It was found that the increase of the tert-butanol/2-butanol ratio affects the liquid-liquid equilibria in the quasi-binary system at ambient pressure in the same way as the increase of pressure modifies the phase behavior of binary 2-butanol–$H_2O$ solutions. The correlation length of critical fluctuations near the liquid-liquid separation and the size of mesoscale droplets of solubilized cyclohexane were obtained by dynamic light scattering and by small-angle neutron scattering. It is shown that the effect of the presence of small amounts of cyclohexane on the near-critical phase behavior is twofold: 1) the transition temperature changes towards increasing the two-phase domain; 2) long-lived mesoscopic inhomogeneities emerge in the macroscopically homogeneous domain. These homogeneities remain unchanged upon approach to the critical point of macroscopic phase separation and do not alter the universal nature of criticality. However, a larger amount of cyclohexane generates additional liquid-liquid phase separation at lower temperatures.

1. Introduction

Hydrotropes are substances consisting of amphiphilic molecules whose non-polar part is much smaller when compared with long hydrophobic chains of traditional surfactants [1,2]. Typical examples of non-ionic hydrotropes include short-chain alcohols and amines. In aqueous environment, hydrotropes unlike surfactants do not spontaneously self-assemble to form stable equilibrium micelles [3], although they are frequently used as "co-surfactants" to stabilize micro-emulsions [4]. However, above a certain concentration, hydrotrope solutions may exhibit dynamic, non-covalent molecular clustering that may be viewed as "micelle-like" or structural fluctuations [5-13]. Molecular dynamics studies show that the clusters of hydrotrope molecules, hydrogen-bonded with the aqueous surrounding, are short-ranged (order of one nm in size) and short-lived with a lifetime of tens or hundreds picoseconds [10-12]. This molecular clustering, usually involving hydrogen bonds with water molecules, can cause distinct thermodynamic anomalies, such as those observed in aqueous alcohol solutions in water-rich region [11,14-19].

In addition, many experiments on aqueous solutions of hydrotropes show the presence of long-lived mesoscale inhomogeneities of the order of a hundred nm in radius [12,13,20-32]. It has been shown that such inhomogeneities occur in aqueous solutions of nonionic hydrotropes when the solution contains a third, more hydrophobic, component [12,13,28-32]. Remarkably, these inhomogeneities (emerging in water-rich ternary systems) are only pronounced in the hydrotrope concentration range where molecular clustering and thermodynamic anomalies are observed in the original binary hydrotrope-water solutions. The hypothesized structure of these mesoscopic droplets is such that they have a hydrophobe-rich core, surrounded by a H-bonded shell of water and hydrotrope molecules [12,13,22,31,32]. These droplets can be extremely long-lived, being occasionally stable for over a year [32]. The phenomenon of the formation of mesoscopic inhomogeneities in aqueous solutions of nonionic hydrotropes induced by hydrophobic "impurities" is referred to as "mesoscale solubilization" [12,13,31,32]. Mesoscale solubilization may represent a ubiquitous feature of certain nonionic hydrotropes that exhibit molecular clustering in water and may have important practical applications in areas, such as drug delivery, if the replacement of traditional surfactants is necessary.

Mesoscale solubilization is still poorly understood and a theory for this phenomenon is yet to be developed. If such a system is brought into the vicinity of the liquid-liquid critical point, mesoscale solubilization could somehow interact with the mesoscopic concentration fluctuations [31]. The first studies of the mesoscale solubilization near the liquid-liquid separation in ternary solutions of *tert*-butanol, propylene oxide and water were performed by Sedlák and Rak [30] and by Subramanian *et al*. [31]. These works agree on the interpretation of the observed mesoscale inhomogeneities as the phenomenon of solubilization of oily impurities originally present in *tert*-butanol and propylene oxide. However, behavior of these inhomogeneities in the critical region remained unclear. Subramanian *et al*. [31] speculated that the mesoscale droplets could exhibit anomalous curvature fluctuations in the critical region, thus forming a sponge phase. Herzig *et al*. [33] have demonstrated that a bicontinuous interface can be stabilized with the help of colloidal particles, which are wetted in both phases. It was thus hypothesized that the mesoscopic droplets, which might behave as colloidal particles, could stabilize the bicontinuous phase formed in hydrotrope-water-oil systems near the critical point. This is one possible scenario. An alternative possibility is the aggregation or disintegration of solubilized droplets caused by attractive Casimir forces induced by the concentration fluctuations [34].

There are two main obstacles preventing quantitative studies of mesoscale solubilization near phase separation and, in particular, in the critical region. Firstly, mesoscale solubilization could be easily confused with the so-called "ouzo effect" [35,36], the formation of micron-size metastable droplets (making the solution milky-white) by nucleation between the thermodynamic solubility limit (binodal) and the absolute stability limit of the homogeneous phase (spinodal). Moreover, mesoscale solubilization and the ouzo effect may overlap, thus making unambiguous interpretations of the experiments challenging. Secondly, in the critical region this task will require simultaneous measurements of two mesoscopic lengths, the correlation length of the critical fluctuations and the size of the mesoscopic droplets. This task cannot be accomplished by light scattering experiments alone. When the correlation length reaches 20-30 nm, the critical opalescence dominates the light-scattering intensity and the size of the mesoscale droplets becomes poorly detectable. Refractive-index matching is not applicable for this particular case because of the large difference between the refractive indices of oil and water. The problem can be solved in small-angle neutron scattering (SANS) experiments by changing the $H_2O/D_2O$ ratio, such that the signal from the SANS scattering caused by the critical fluctuations can be almost eliminated.

In this work we report the results of dynamic light scattering (DLS) and SANS experiments on binary solutions, 2,6-lutidine–$H_2O$, and ternary (quasi-binary) solutions, 2,6-lutidine–$H_2O/D_2O$ and *tert*-butanol/2-butanol–$H_2O$. All these systems contain hydrotropes: 2,6-lutidine, *tert*-butanol (TBA), and 2-butanol (2BA), and all exhibit critical phase separation. Critical phenomena in aqueous solutions of 2,6-lutidine were studied earlier in detail [37-42]. Liquid phase separation in this system was also studied in the presence of colloid particles [34,43]. In particular, attractive Casimir forces between colloid particles, induced by the concentration fluctuations, were discovered for this system [34].

The systems 2,6-lutidine–$H_2O/D_2O$ and TBA/2BA–$H_2O$ can be considered as a "quasi-binary". Mixtures of isotopes, $H_2O/D_2O$, as well as of isomers, TBA/2BA, form near-ideal solution and can be regarded as effective compounds whose properties can be tuned by changing the component ratio. In particular, the two butanol isomers are very similar, except for the fact that 2BA exhibits partial immiscibility with water, while TBA is the highest molecular-weight alcohol to be completely miscible with water under ambient conditions. TBA is a "perfect" hydrotrope. When placed at the water/oil interface, a TBA molecule is equally divided, with the hydroxyl group on the water side and the methyl groups on the oil side [44]. It was expected that on addition of TBA to 2BA the immiscibility gap in the TBA/2BA–$H_2O$ system would shrink and a liquid-liquid critical point could emerge. We have confirmed this expectation. Moreover, we have shown that an increase of the TBA/2BA ratio affects the liquid-liquid equilibria in the quasi-binary system at ambient pressure in the same way as the increase of pressure modifies the phase behavior in binary aqueous solutions of 2BA [45-47].

We were able to simultaneously measure the correlation length of the critical fluctuations near the liquid-liquid separation and the size of the mesoscopic inhomogeneities. It is shown that the effect of the presence of small amounts of a hydrophobe (controlled addition of cyclohexane) on the near-critical phase behavior is twofold: 1) increase of the domain of liquid-liquid separation; 2) formation of long-lived mesoscopic inhomogeneities that survive and remain basically unchanged near the critical point of macroscopic phase separation. The presence of these inhomogeneities does not alter the universal nature of critical anomalies. However, larger amounts of cyclohexane generate additional macroscopic liquid-liquid phase separation at lower temperatures.

## 2. Experimental Methods

### 2.1 Materials

TBA (packaged under argon) with a labeled purity of 0.998 was purchased from Alfa Aesar and 2BA from Sigma-Adrich with a labeled purity of 0.995+. Cyclohexane (CHX) with a labeled purity of 0.99+ was purchased from Merck (used in Moscow) and from Baker, 0.999 (used in Maryland). We used deionized dust-free Milli-Q water (18.2 MΩ·cm) in Maryland and water for medical injections (r-phar.com) in Moscow. Deuterium oxide ($D_2O$) was obtained from Cambridge Isotopes, 99.9% D. In two labs we used two different sources of 2,6-lutidine, one (Maryland) was from Aldrich, 99%+, and another one (Moscow) from Merck with a labeled purity of 0.98+.

We used two different protocols of adding cyclohexane into the solutions of hydrotropes. At NIST, the samples with controlled amounts of cyclohexane were prepared by first mixing CHX and 2,6-lutidine and then by rapidly adding the water. For the NIST DLS experiments, the components were filtered with 200 nm Nylon filters to remove dust particles and mixed. In another protocol, CHX was added to homogeneous aqueous solutions of 2,6-lutidine (Moscow) and TBA/2BA (College Park). The samples with TBA/2BA–$H_2O$ solutions were prepared by adding 2BA, then TBA, and then water to a vial through a 20 nm Anopore filter.

Aqueous solutions of 2,6-lutidine and TBA/2BA did not show the presence of mesoscale inhomogeneities prior addition of CHX, when tested by DLS at the scattering angle of 90 degrees. However, in solutions of 2,6-lutidine (labeled purity 0.98+) and in solutions of TBA/2BA a signal from 100-200 nm droplets, supposedly caused by the presence of hydrophobic impurities, was detected at small scattering angles, 30 and 45 degrees, even before addition of CHX. However, based on the study of the effect of CHX addition on the critical temperature, we have concluded that the level of the hydrophobic impurities originally present in these systems does not exceed ~0.1%.

## 2.2   *Supplemental Information*

To extract the size of Brownian particles and the correlation length of critical fluctuations from DLS experiments one needs the information on the refractive index and shear viscosity. The refractive index enters the definition of the light-scattering wave number and the viscosity is involved in the expression for the diffusion coefficient.

The refractive index was measured with an Abbe refractometer at room temperature for TBA/2BA–$H_2O$ solutions or taken from the literature (for 2,6-lutidine–water [48]). The refractive index, *n*, at 22 °C for TBA/2BA solution, with 59.69 % (mass) $H_2O$, 33.55 % 2BA, and 6.75 % TBA that exhibits a critical consolute point at atmospheric pressure, was found to be *n* = 1.3685. Since the temperature interval in our measurements was relatively narrow and since the estimated effect is comparable with typical errors in the determination of the diffusion coefficient, the temperature dependence of the refractive index was neglected. The viscosity of the solutions 2,6-lutidine–water is known from the literature for 30 % mass of lutidine for the temperature range studied [49] and at 22 °C for 15, 45 and 60 % [48]. The values of viscosity for these concentrations at other temperatures were interpolated.

To obtain the viscosity of the TBA/2BA solution, with 59.69 % (mass) $H_2O$, 33.55 % 2BA, and 6.75 % TBA that exhibits a critical consolute point at atmospheric pressure, calibrated latex particles with a given radius of 50 nm were suspended in solution. The viscosity was calculated via the Stokes-Einstein equation that relates the Brownian diffusion, measured by DLS from the latex particles, and the hydrodynamic radius of the particles (see Section 2.4, Eq. 3). The viscosity data and their linear regression are shown in Figure 1. It was difficult to measure the viscosity at higher temperatures, closer to the critical temperature (23.43±0.02 ºC), because the critical opalescence overwhelmed the scattering from the latex particles. This is why the viscosity of this solution at higher temperatures (up to 23.4 °C) was obtained by linear extrapolation.

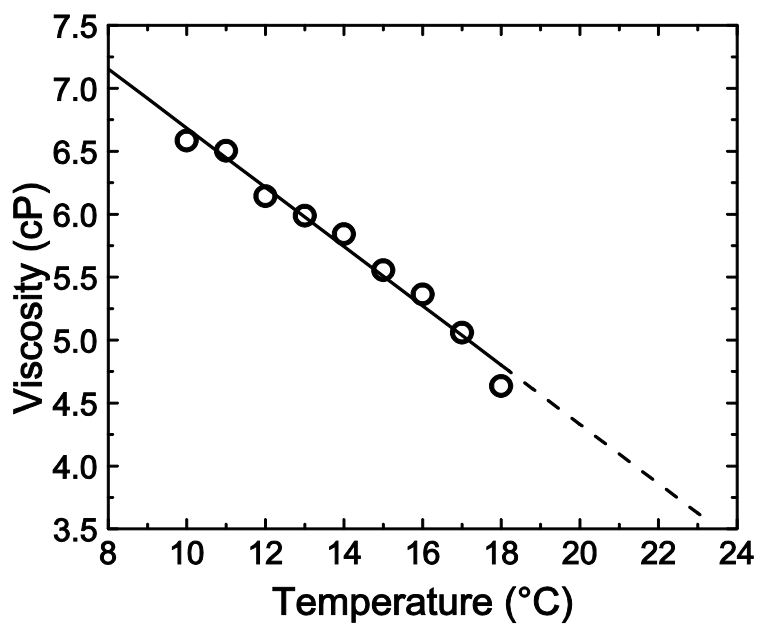

Figure 1. Shear viscosity as a function of temperature for the ternary solution of 59.69 % (mass) H$_2$O, 33.55 % 2BA, and 6.75 % TBA measured by DLS from calibrated latex particles.

*2.3   Determination of Phase Behavior*

The liquid-liquid phase transition temperatures in 2,6-lutidine–water upon addition of cyclohexane were detected visually and from the location of a maximum in the light-scattering intensity. Second macroscopic phase separation at lower temperatures was detected upon addition of CHX more than 0.16 % (mass). Ternary phase diagram (H$_2$O–TBA–2BA) was determined by the cloud-point method [42]. The third component (2BA) was added to a binary mixture in small steps. At each step, the ternary mixture was manually shaken and let to rest for about 3 to 5 minutes. The sample was then visually observed to determine if phase separation had occurred. If not, more of the third component was added and the above procedure was repeated. The ternary phase diagram of the TBA/2BA–water system was determined at room temperature, 22 °C ± 0.5 °C. In order to estimate the location of the critical point, light scattering experiments were carried out in the macroscopic one-phase region close to the binodal curve. If the correlation length of critical fluctuations exhibited a maximum, then the point of the binodal curve corresponding to this maximum was interpreted as the solution with the critical composition.

## 2.4 Light Scattering

Static and dynamic light scattering experiments were performed in College Park and Moscow with Photocor setups, as described in refs. [28,29,31,32]. In Moscow, the measurements were performed at the scattering angles of 20, 30 and 90 degrees. Measurements in College Park were performed from 30 to 130 degrees in 10-degree increments. Light scattering measurements at NIST were made at 90 degree with a Wyatt DynaPro NanoStar setup. Temperature was controlled with an accuracy of ± 0.02-0.05 °C that prevented us from performing experiments in the immediate vicinity of the critical point.

For two exponentially decaying relaxation processes, the intensity auto-correlation function $g_2(t)$ (obtained in the homodyning mode) is given by [50,51]

$$g_2(t) - 1 = \left[ A_1 \exp(-t/\tau_1) + A_2 \exp(-t/\tau_2) \right]^2, \qquad (1)$$

where $A_1$ and $A_2$ are the amplitudes of the two relaxation modes, $t$ is the "lag" (or "delay") time of the photon correlations and $\tau_1, \tau_2$ are the characteristic relaxation times. For two diffusive relaxation processes, the relaxation times are related to the diffusion coefficient $D_1$ and $D_2$ as [50,51]

$$\tau_{1,2} = \frac{1}{D_{1,2} q^2}, \qquad (2)$$

where $q$ is the difference in the wave number between incident and scattered light, $q = (4\pi n/\lambda)\sin(\theta/2)$, $n$ is the refractive index of the solvent, $\lambda$ is the wavelength of the incident light in vacuum ($\lambda$ = 633 nm for He-Ne laser) and $\theta$ is the scattering angle. The linear dependence between the decay rate $1/\tau$ and $q^2$ is characteristic for diffusive relaxation [50,51]. For mono-disperse, spherical Brownian particles the hydrodynamic radius $R$ can be calculated with the Stokes-Einstein relation [50]:

$$R = \frac{k_B T}{6\pi \eta D_2}, \qquad (3)$$

where $D_2$ is the diffusion coefficient of the particles, $k_B$ is Boltzmann's constant, $T$ is the temperature and $\eta$ is the shear viscosity of the medium.

The correlation length of critical fluctuations, $\xi$, was obtained from the mutual diffusion coefficient $D_1$ which near the critical point was approximated as:

$$D_1 = \frac{k_B T}{6\pi\eta\xi}\left[K(q\xi) + \frac{\xi_0}{\xi}\right], \qquad (4)$$

where $K(q\xi) = \frac{3}{4}\{1+(q\xi)^{-2} + [q\xi - (q\xi)^{-3}]\tan^{-1}(q\xi)\}$ is the Kawasaki function [52], $\xi_0$ is the "bare" correlation length (order of the range of molecular interactions). Eq. (4) is a simplified version of the mode-coupling theory [52-56]. In this version, the weak critical singularity of shear viscosity [55,56] and a $q$-dependence of the non-asymptotic term ($\xi_0/\xi$) [57] are neglected. In the temperature interval studied in this work, both these simplifications are reasonable. The experiments were not performed in the immediate vicinity of the critical point (closer than $T_c - T \cong 0.1°C$), where the anomaly of the viscosity could become significant, while the condition $\xi \gg \xi_0$ was always satisfied. Hence the non-asymptotic term $\xi_0/\xi$ gives a minor correction to the diffusion coefficient in the temperature range studied. At $q\xi < 1$, the Kawasaki function can be approximated as $K(q\xi) \cong 1 + \frac{3}{5}(q\xi)^2$. Since in our DLS experiments, the condition $q\xi < 1$ was always satisfied, the Kawasaki correction affected only a few (closest to the critical temperature) measurements of the diffusion coefficient.

## 2.5   *Small-angle Neutron Scattering*

DLS measurements cannot be used to study the nature of the droplets of solubilized hydrophobe in the immediate vicinity of the critical point because the scattering from the critical fluctuations overwhelms the signal from the droplets. To assess the region where the correlation length becomes comparable to the droplet radius, we used small-angle neutron scattering (SANS) with contrast-matched samples [58]. By using this technique, one can eliminate the contribution from the critical concentration fluctuations and focus on the scattering from other inhomogeneities near the solution critical point. The intensity of SANS depends on differences between scattering-length densities of the constituent molecules. The scattering-length density of 2,6-lutidine is $\rho = 1.16 \cdot 10^{-8}$ nm$^{-2}$, whereas the scattering-length densities of ordinary and heavy water are $\rho = -0.56 \cdot 10^{-8}$ nm$^2$ for H$_2$O and $\rho = 6.40 \cdot 10^{-8}$ nm$^2$ for D$_2$O. Because the scattering-length densities of ordinary and heavy water have opposite signs, an appropriate admixture of H$_2$O and D$_2$O can create a solvent with an average scattering-length density intermediate to these two extreme values. This is the basis of the contrast variation technique. For a 0.7326 mass ratio of H$_2$O and D$_2$O, the average scattering length density matches that of 2,6-lutidine. As a result, the solvent fluctuations do not contribute to the scattering intensity and the scattering from the cyclohexane droplets is isolated. Small-angle neutron scattering measurements were performed at the NIST Center for Neutron Research on the NGB 30 m SANS instrument using a $\lambda = 0.6$ nm incident neutron wavelength. The SANS instrument wave number is defined as $Q = (4\pi/\lambda)\sin(\theta/2)$, where $\theta$ is the scattering angle. By measuring the SANS intensity as a function of the wave number, $I(Q)$ one can obtain the average radius, $R$, of inhomogeneities if their scattering-length density different from that of solvent.

The expression for fitting a spherical droplet model to the SANS intensity data reads [58]:

$$I(Q) = (\Delta\rho)^2 \phi V \left[\frac{3 j_1(QR)}{QR}\right]^2, \qquad (5)$$

with

$$j_1(QR) = \frac{\sin(QR) - QR\cos(QR)}{(QR)^2}, \qquad (6)$$

In Eq. 5, $\Delta\rho$ is the difference in scattering-length densities of the solvent and spherical droplets, $\phi$ is the volume fraction of droplets, and $V = \frac{4}{3}\pi R^3$ is the droplet volume. The droplet radius $R$, obtained from SANS experiments with using Eq. 5, is assumed to be close to the hydrodynamic radius obtained by DLS.

## 3. Results

### 3.1 Phase Behavior

#### 3.1.1 2,6-lutidine–water

Figure 2 demonstrates the phase behavior of 2,6-lutidine solutions. The liquid-liquid coexistence is shown as reported by Grattoni *et al*. [48] who indicated the critical temperature as 34.1 °C and the critical concentration as 28 % (mass) lutidine. However, the location of the critical point is strongly affected by impurities and by the method of determination and varies in the literature between 33 °C and 35 °C for the critical temperature and between 28 % and 30 % for the critical concentration [37-42]. The value of the critical concentration is especially uncertain and the properties are rather insensitive to small variations of the composition near the critical point because of the flat shape of the coexistence curve. At NIST, experiments were performed on the sample containing 29 % lutidine. In experiments performed in Moscow, the sample with 30 % lutidine was considered as the solution with near-critical composition. The critical temperature, defined as the temperature of the divergence of the correlation length, was found to be 34.29 ± 0.02 °C.

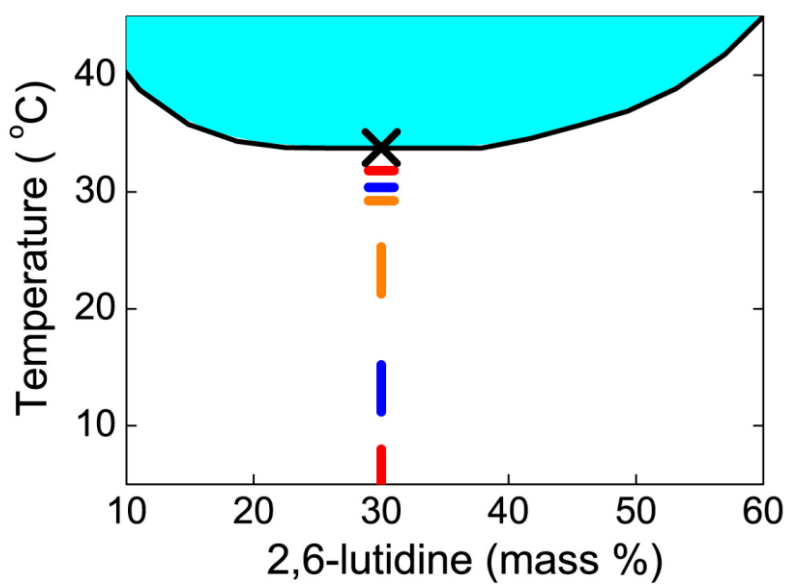

Figure 2. Effects of addition of cyclohexane on the liquid-liquid separation in 2,6-lutidin aqueous solutions at 30 % (mass) lutidine. Solid curve is the coexistence curve before the addition of CHX. Cross is the critical point. Three upper horizontal bars show the temperatures of the critical phase separation (obtained upon heating) shifted by addition of cyclohexane (red is 0.16 % (mass) CHX,

blue 0.32 %, and orange 0.41 %). Three vertical bars at lower temperatures show the locations of first-order phase separation smeared over several ºC (obtained upon cooling with a rate of 3 ºC /hour). As for the upper (horizontal) bars, for these bars red is 0.16 % (mass) CHX, blue 0.32 %, and orange 0.41 %.

Addition of CHX, as expected, increases the two-phase domain in 2,6-lutidine solutions by lowering the critical temperature. A possible change in the critical concentration is within the accuracy of its determination. The addition of 0.1 mass % CHX lowers the critical temperature by 1°C as illustrated by Figure 3a. However, a remarkable effect of additions of CHX is the emergence of the second liquid-liquid separation at lower temperatures. The second liquid-liquid transition was detected by a sharp increase of the light-scattering intensity (shown in Figure 3b) and by visual observations of sample's behavior (Figure 4). First the solution is emulsified and then phase-separated upon further cooling with the light-scattering intensity decreasing. This is the first-order transition: the peak in the intensity is attributed to the formation and growth of droplets of the new phase by nucleation between the equilibrium transition point and the limit of stability (spinodal). The solution becomes milky-white. This spectacular phenomenon is known as "ouzo effect" [35,36] which is controlled by kinetics of nucleation. The apparent width of the transition (up to several °C) depends on the cooling rate, however the transition is reproducible.

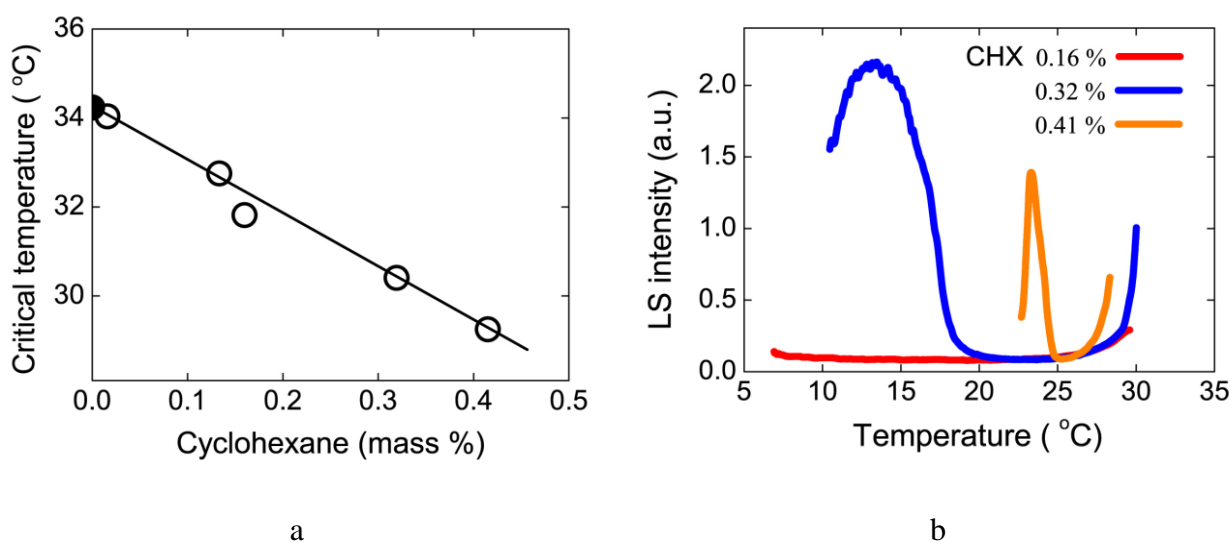

a                                           b

Figure 3. Determination of liquid-liquid separation in 2,6-lutidine–water upon addition of cyclohexane. (a) Shift of the critical temperature. Closed semicircle is the critical temperature without CHX. (b) Light-scattering intensity upon cooling (at the rate of 3 ºC/hour) for the samples with different additions of CHX.

Upon heating and equilibrating for 1.5 hour, the sample becomes clear and is phase-separated above the critical temperature. The phenomenon is reproducible: the ouzo effect comes back upon cooling, as shown in Figure 4. However, after equilibrating for a few (up to 24) hours, the solution becomes clear again because the nucleation is followed by Ostwald ripening with the ultimate formation of a thin layer of cyclohexane on the top.

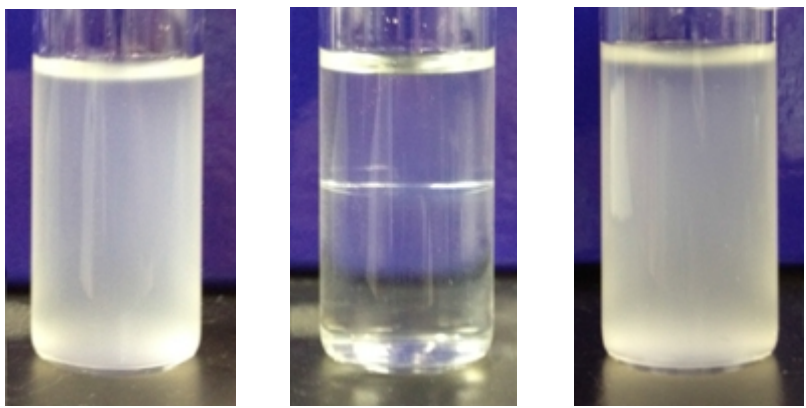

Figure 4. Ouzo effect in the solution of 2,6-lutidine (29 %)–water with 0.5 % cyclohexane. From the left to the right: the sample prepared by adding the water at room temperature; the sample heated up to 30 ºC (above the critical demixing temperature of this sample); the sample is cooled back to the room temperature.

The phase behavior in the quasi-binary mixture 2,6-lutidine–$H_2O/D_2O$ is very similar to that of 2,6-lutidine–$H_2O$. However, the critical temperature for the sample containing 73.26/26.74 % ratio of $H_2O/D_2O$ is lower. For the binary solution with 0.2 % CHX, the critical temperature was found to be around 31.5-31.6 °C.

*3.1.2 TBA/2BA–water*

Figure 5 demonstrates the phase behavior of the ternary system TBA–2BA–$H_2O$ at 22°C. For this temperature the critical composition was found to be about 59.70 % (mass) $H_2O$, 33.55 % 2BA, and 6.75 % TBA. In a ternary system, at constant pressure, there is a critical line ($T_c$ is a function of the ternary composition). Therefore, the critical point can be reached by further tuning the temperature of the sample which composition is close to critical. For this particular sample, the critical temperture was found to be 23.41 ±0.02 °C. Addition of 0.4 mass % CHX lowered the critical temperature by 4.3°C, similar to the effect of adding CHX on the critical temperature of lutidine–water solution, as shown in Figure 3a.

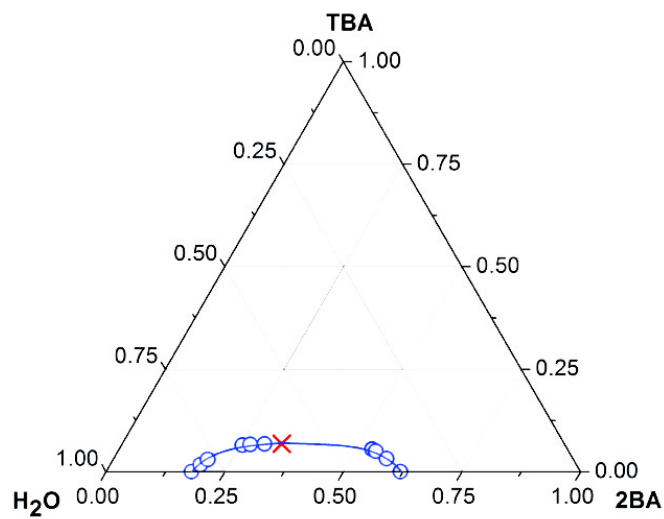

Figure 5. Ternary phase diagram of TBA–2BA– $H_2O$ at room temperature (22 ºC). Composition is shown in mass fraction. Area below the blue curve is the two-phase region. Circles are experimental data. Cross is the estimated location of the critical point.

Figure 6 demonstrates that the effect of addition of TBA into 2BA–$H_2O$ solutions is similar to the effect of increasing the pressure in the binary solution 2BA–$H_2O$. Experimental data [45,46] show that this binary solution at 200 atm exhibits two critical points (at about 25 ºC and below 0 ºC). For the quasi-binary TBA/2BA–$H_2O$ solution, similar behavior, based on the observed location of the critical point, can be projected.

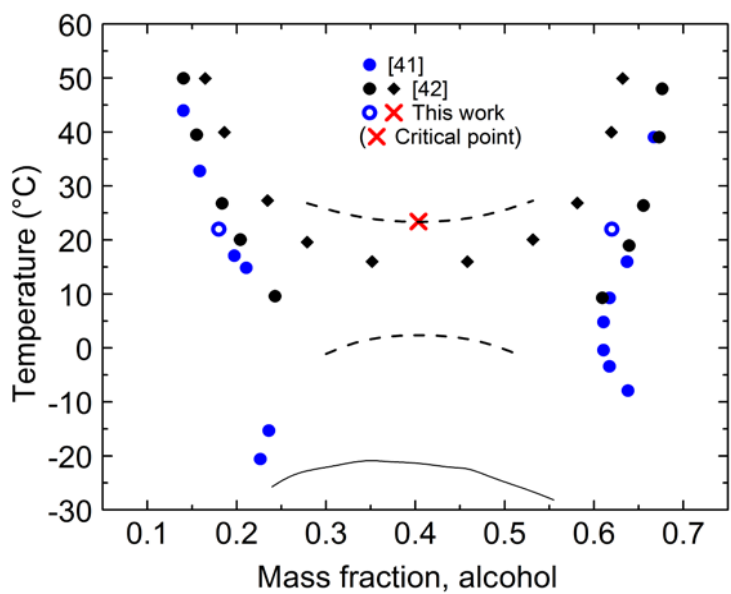

Figure 6. Miscibility gap in the binary system 2BA–water (closed circles are experimental data [45,46]) and projected two braches (upper and lower, shown by dash) of near-critical phase coexistence in the quasibinary system with 59.70 % (mass) $H_2O$, 33.55 % 2BA, and 6.75 % TBA at 1 atm (about 40 % alcohol). Cross is the critical point at the upper branch. Black diamonds are data at 200 atm [46]. Dashed curves is a qualitative prediction of upper and lower phase separation in the quasibinary system TBA/2BA–water at 1 atm. Solid curve is modeling of the binary 2BA– $H_2O$ system [47].

## 3.2 Effects of addition of cyclohexane on mesoscale behavior of 2,6-lutidine in $H_2O/D_2O$ contrast-matched quasi-binary mixture (DLS and SANS)

Figure 7a displays the DLS autocorrelation functions at the scattering angle of 90 degrees in a quasi-binary system 2,6-lutidine–$H_2O/D_2O$ at 20 °C for three cyclohexane concentrations (in mass %). For the largest concentration (0.57 %) the correlation function is a single exponential. Cyclohexane droplets dominate the scattering (apparently by the developed ouzo effect), while the solvent fluctuations are barely distinguishable. Both the scattering from mesoscale inhomogeneities and the concentration fluctuations clearly contribute at 0.2 % CHX - the correlation function is a double exponential. This sample is apparently clear at room temperature, thus the mesoscale inhomogeneities can hardly be attributed to the ouzo effect. Finally, at 0.12 % CHX the scattering at the angle of 90 degree from the mesoscale inhomogeneities is weak and is dominated by the scattering from the concentration fluctuations: the correlation function is again a single exponential. The intensity of scattering from spherical particles is proportional to both the particle number density and the size of the particle. In our case, not only the volume fraction of the cyclohexane is changing, the size of the particles is also changing. This explains the drastic increase in the droplet scattering contribution as the cyclohexane volume fraction is increased. Applying the Stokes-Einstein relationship to the cyclohexane droplets, we find that the average radius of the mesoscale inhomogeneities is $R \approx 70\text{-}80$ nm for 0.2 % CHX (mesoscale solubilization) and $R \approx 300\text{-}400$ nm for 0.57 % CHX. The visual observations of 2,6-lutidine–$H_2O$ solutions (as demonstrated in Figure 4) suggest that the studied quasibinary 2,6-lutidine–$H_2O/D_2O$ solution with 0.57 % CHX at 20 °C are most likely in the stage of developing phase separation and the radius 300-400 nm corresponds to metastable droplets formed by nucleation of a new phase (ouzo effect).

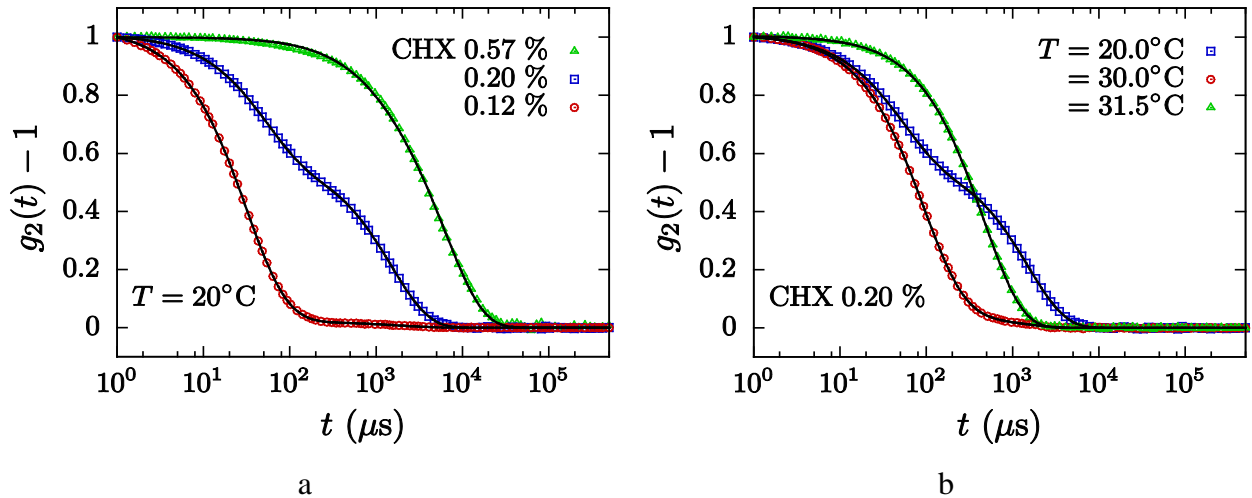

Figure 7. DLS correlation functions for the quasi-binary system 2,6-lutidine (29 %)–H$_2$O/D$_2$O (left) addition of different amounts of cyclohexane (0.12, 0.2, and 0.57 mass % of CHX) at 20 ºC; (right) Effect of temperature on the shape of the correlation function with 0.2 mass % CHX. Solid curves are fits to Eq. 1.

The effect of temperature on the scattering from the 0.2 % CHX sample is presented in Figure 7b. At low temperature ($T$ = 20 °C) both the contributions from the mesoscale-droplet scattering and the concentration-fluctuation scattering can be clearly distinguished. At 30 °C the magnitude of concentration fluctuations has already completely buried the contribution from droplet scattering. As the temperature increases further (31.5 °C), the correlation function exhibits the characteristic slowing-down of critical fluctuations and the correlation length of the fluctuations approaches the droplet size. From these measurements we can infer that dynamic light scattering measurements cannot be used to study the nature of the mesoscopic cyclohexane droplets in the close vicinity of the critical point because the scattering from the critical fluctuations overwhelms scattering from the droplets. The correlation lengths calculated (with Eqs. 2 and 4) from the correlation functions, presented in Fig. 7b, turned to be 1.9 nm at 20°C, 5.2 nm at 30°C, and 38.5 nm at 31.5°C.

The results from the contrast-matched SANS measurements are shown in Figure 8 for two temperatures for the sample with 0.2 % CHX. Although the data suffer from a fair degree of statistical uncertainty, the mesoscopic droplets can be clearly distinguished. From a fit to a mono-disperse spherical form-factor model we find that the average radius of the droplets to be $R$ = 77 nm.

This value is quite close to the effective hydrodynamic radius (~80 nm) measured by DLS.

We did not observe any detectable change in the SANS intensity data for this sample as the critical point was approached. This fact also confirms that the 2,6-lutidine and the solvent ($H_2O/D_2O$ mixture) have been effectively contrast-matched.

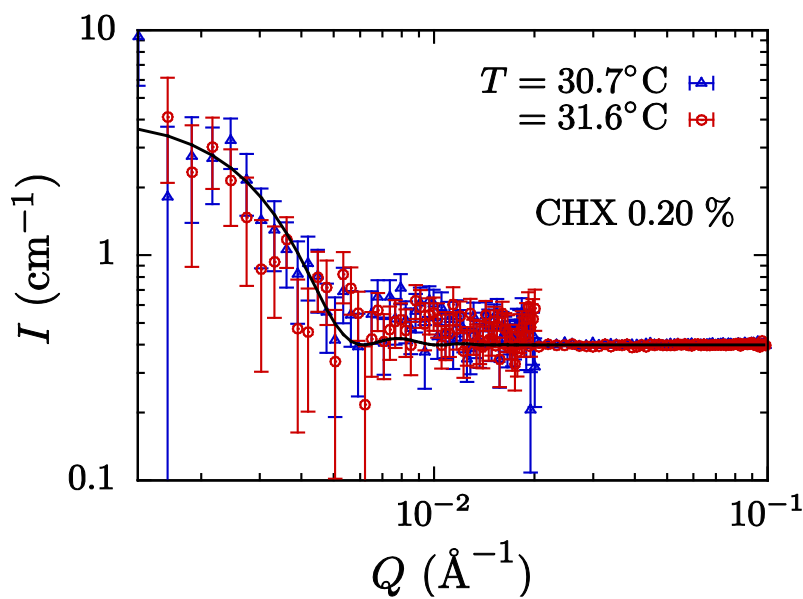

Figure 8. SANS intensity as a function of the wave number in the quasi-binary system 2,6-lutidine (29 %)– $H_2O$ / $D_2O$ at two different temperatures, 30.7°C and 31.6°C for 0.2 mass % CHX. Solid line is fit of a mono-disperse spherical form-factor model to the data at 30.7°C by Eqs. 5 and 6.

We want to emphasize that the samples used in the SANS measurements and in the DLS measurements presented in Fig. 7 were identical. Hence, we conclude that the critical fluctuations do not have a significant impact on the mesoscopically solubilized cyclohexane, at least if the system is not extremely close to the critical point.

There is an important feature of the behavior of mesoscale droplets of solubilized hydrophobic impurities in solutions of hydrotropes, which was reported in the literature [28-32]. While the droplets are extremely long-lived, this may be a non-equilibrium phenomenon. In particular, the droplet size depends on the protocol used to prepare the sample (aging and cooling/heating rates [29]). On multiple occasions, we had observed changes in the droplet size from 80-90 nm to 120-130 nm after a day of waiting and re-measuring, as shown in Figure 9, while the contribution from the concentration fluctuations remained unchanged.

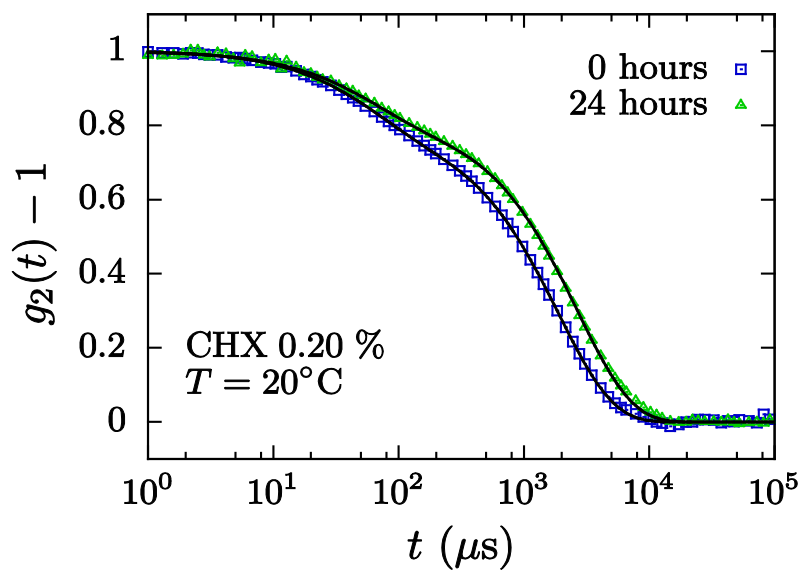

Figure 9. Effect of aging on the size of mesoscopically solubilized cyclohexane droplets at 20 ºC in the quasi-binary system 2,6-lutidine (29 %)– $H_2O$ / $D_2O$ with 0.2 mass % CHX. It is seen that the fast mode (relaxation of the concentration fluctuations) remains unchanged.

### *3.3 Effects of addition of cyclohexane on mesoscale and phase behavior of 2,6-lutidine in water (DLS)*

Long-term stability and aging of mesoscale droplets was investigated in more detail for solutions of 2,6-lutidine in ordinary water ($H_2O$) with addition of cyclohexane. The results of monitoring (by DLS, at a scattering angle of 30 degree and at room temperature) two samples, 30 % (critical) and 15 % lutidine, with addition of 0.02 mass % CHX are shown in Figure 10. The size of the mesoscale droplets for the 30 % sample (Figure 11a) was stable for many days of continuous monitoring. Monitoring of two other samples, 45 % and 60 % lutidine, produced the results very similar to those for the 30 % sample.

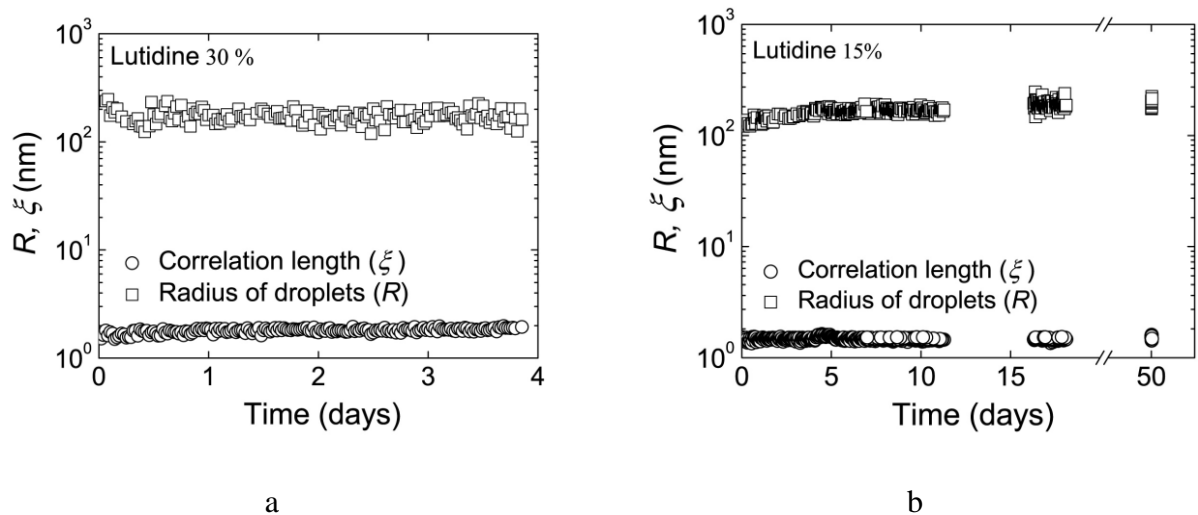

a                      b

Figure 10. Long-term monitoring (at 25°C and at the scattering angle of 30 degrees) the measurements of two characteristic length-scales (the correlation length of concentration fluctuations and the radius of mesoscopic droplets) in 2,6-lutidine–water solutions with addition of 0.02 % (mass) cyclohexane. (a) 30 % (mass) 2,6-lutidine; (b) 15 % (mass) 2,6-lutidine.

However, a longer monitoring the 15% sample (Figure 10b) revealed ripening of the droplets from about 100 nm to about 200 nm, observed during the first five days of monitoring. Then the size was stabilized. The effect of ripening was not observed for samples with larger content of lutidine.

The average size of the mesoscale droplets also depends on the lutidine/water ratio. As shown in Figure 11, the size increases with decrease of the concentration of lutidine. This effect is obviously linked to increasing hydrophobicity of the system with decrease of the lutidine concentration. Figure 12 demonstrates how this size also increases with increase of the amount of added CHX, as well as upon cooling.

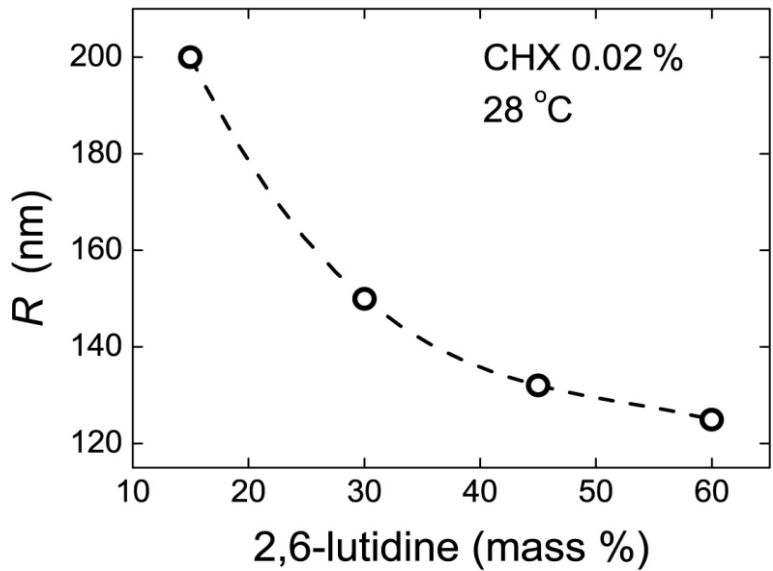

Figure 11. The size of mesoscopic droplets (DLS at 30-degree scattering angle) in aqueous binary solutions of 2,6-lutidine with addition of 0.02 % CHX as a function of lutidine concentration.

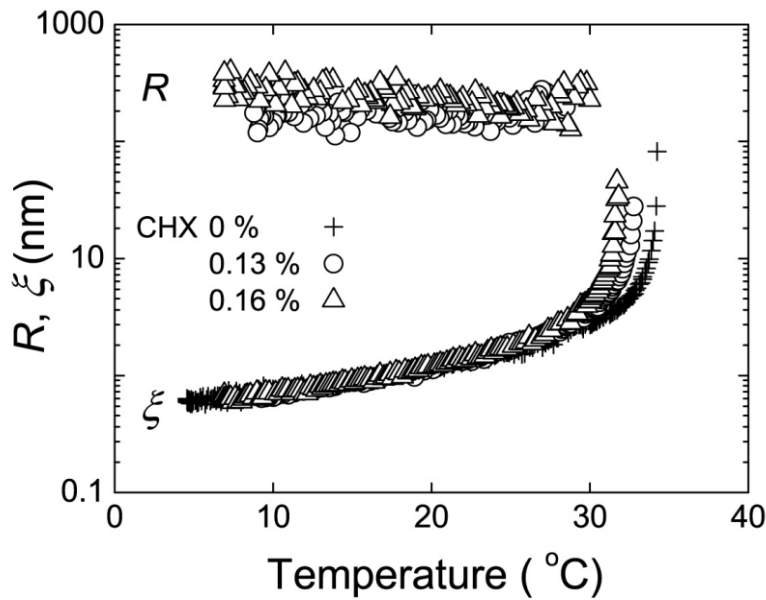

Figure 12. The size of mesoscopic droplets in aqueous binary solutions of 2,6-lutidine with addition of 0.13 and 0.16 % CHX and the correlation length of critical fluctuations vs. temperature for solutions without CHX, 0.13 %, and 0.16 % CHX, measured by DLS at 30-degree scattering angle.

In Figure 12 we also show the growth of the correlation length of critical concentration fluctuations (obtained from the diffusion coefficient given by Eq. 4) for three samples (without CHX, 0.13 %, and 0.16 % CHX). When the correlation length becomes comparable with the size of the mesoscale droplets, the scattering from critical fluctuations overwhelms the scattering from the droplets and the droplet size becomes undetectable.

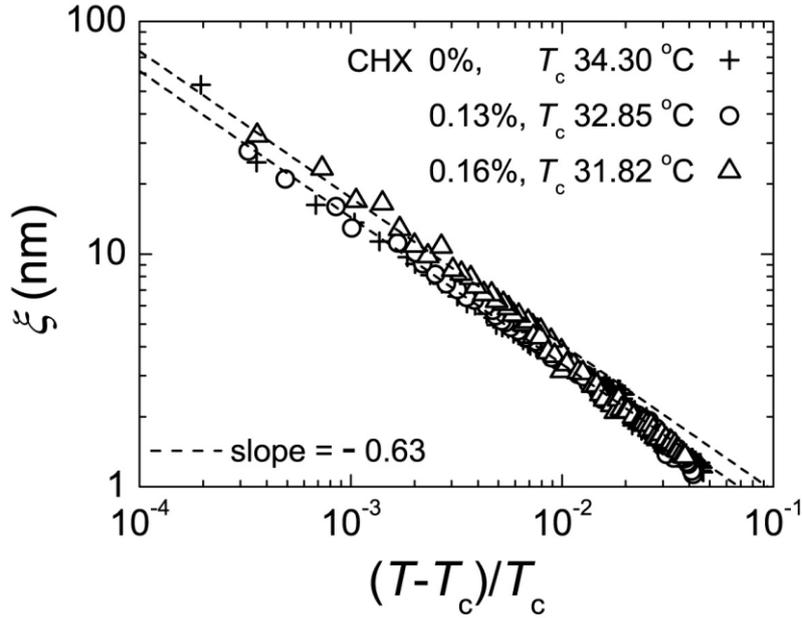

Figure 13. Correlation length in the critical mixture 2,6-lutidine–water. The slope of two dashed lines corresponds to the universal critical exponent. The bare correlation length ($\xi_0$) is 0.185 nm (for the solutions without cyclohexane and with 0.13 % CHX) and is 0.225 nm for 0.16 % CHX.

In Figure 13 the correlation length of critical fluctuations as a function to proximity to the critical temperature is presented in log-log scale. In the interval $(T-T_c)/T_c < 10^{-2}$ the data for the solution without CHX and with 0.13 % CHX can be well described by a scaling power law as

$$\xi = \xi_0 \left(\frac{T-T_c}{T_c}\right)^{\nu}, \qquad (7)$$

with the universal critical exponent (fixed) $\nu = 0.63$ and $\xi_0 = 0.185$ nm. This result is in agreement with previous studies of criticality in this system, $\xi_0 = 0.2 \pm 0.02$ nm [38]. The data for the solution with 0.16 % CHX systematically deviate from the data obtained for the other solutions, however, within the statistical error they are still consistent with the power law given by Eq. 7. We also notice that with increase of the CHX concentration the bare correlation length increases.

## 3.4 Critical phenomena in a quasi-binary system 2BA/TBA – water

In Figure 14 we show the coefficient of mutual diffusion in the near-critical quasibinary solution of TBA/2BA in water, measured by DLS at different scattering angles, as a function of temperature. The solid curve is a fit of a semi-empirical formula, given as

$$D = A(T_c - T)^\nu (T - T^*)^\nu + B(T_c - T)(T - T^*), \qquad (8)$$

where $T_c$ (= 23.41 °C) is the critical temperature, $\nu = 0.63$ is the critical exponent of the correlation length that provides the diffusion coefficient to vanish at the critical point. $T^*$ is projected ("virtual") temperature of possible vanishing the diffusion coefficient at a lower temperature, as suggested by the phase diagram presented in Figure 5. This formula is a further simplified version of Eq. 4 in which the Kawasaki function is neglected and the non-asymptotic term as well as the temperature dependence of viscosity are absorbed by the second (empirical) term. The asymptotic scaling power law, given by Eq. 7, approximates the correlation length diverging at the critical point and is applied twice: providing vanishing diffusion coefficient at the experimentally observed critical temperature and at a lower (projected) temperature $T^*$.

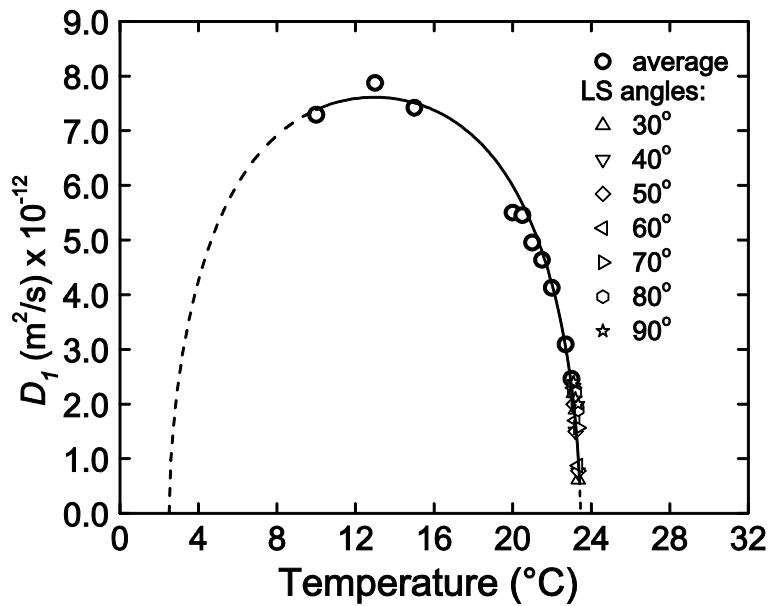

Figure 14. Coefficient of mutual diffusion in the quasi-binary solution containing 59.70 % (mass) H$_2$O, 33.55 % 2BA, and 6.75 % TBA. The left dashed line is a qualitative guidance, being the result of extrapolation by semi-empirical Eq. 8.

The diffusion coefficient becomes wave-number/scattering-angle dependent in the close vicinity of the critical temperature. This effect is accounted by the Kawasaki function $K(q\xi)$ in Eq. 4. For the temperature further away from the critical point, where $K(q\xi) = 1$, the diffusion data, plotted in Figure 14, are averaged for 11 scattering angles, from 30 to 130 degree.

The correlation length, calculated from the experimental data on mutual diffusion with full implementation of Eq. 4, is presented in Figure 15. It is demonstrated that the correlation length within the statistical error (however relatively large when compared with the results presented in Figure 13) is consistent with the same scaling power law. Upon departure from the critical point the correlation length deviates from the power law and saturates at about 10-15 °C.

The only significant difference between these results and those presented in Figure 13, is the difference in the bare correlation length. For the quasibinary solution TBA/2BA–H$_2$O this length is 0.45 nm, twice larger that for solutions of lutidine.

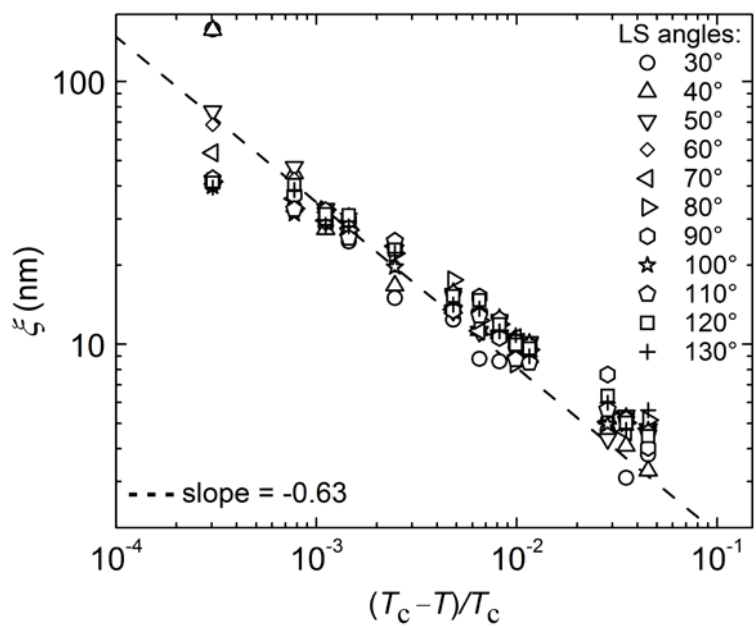

Figure 15. Correlation length in the quasi-binary solution TBA/2BA–H$_2$O containing 59.70 % (mass) H$_2$O, 33.55 % 2BA, and 6.75 % TBA. The critical temperature is fixed at $T_c = 23.43\,°C$.

**4.     Discussion**

Addition of small amounts of cyclohexane (hydrophobe) lowers the critical consolute temperature of the 2,6-lutidine–water system, as expected, and generates a second liquid-liquid separation at lower temperatures, as can also be easily understood. With an increase of hydrotrope concentration the second liquid-liquid separation moves to higher temperatures and would overlap the original immiscibility in the 2,6-lutidine– $H_2O$ system that is moving to lower temperatures upon addition of hydrotrope. This overlapping generates three-phase coexistence. Interestingly, addition of tertiary butanol to the solutions of 2-butanol in water mirrors this phenomenon: initially 2-butanol is partially immiscible with water, but TBA shrinks the miscibility gap and at a certain concentration of TBA two liquid-liquid separation domains (at higher and lower temperatures) emerge. A similar effect can be achieved by applying pressure on the 2,6-lutidine–$H_2O$ system.

However, the major challenge is to fully understand the role and thermodynamic nature of mesoscopic droplets that are formed upon addition of small amounts of a hydrophobe in a broad range of temperatures and concentration of hydrotrope solutions. It seems that the macroscopically homogeneous phase (below the critical temperature but above the temperature of the second macroscopic liquid-liquid separation) behaves as a dilute colloid solution; changing the critical temperature in the same manner as usual colloids do [59], but does not alter the universal nature of the critical behavior. When the solution is macroscopically phase-separated, the mesoscale droplets are observed in the aqueous phase for days or even weeks, however, upon a longer period of time the phase is cleared. Occasionally, a soap-like residual mesophase on the water-oil interphase is observed over the years [32,60]. We were not able to observe effects of critical fluctuations on the mesoscopic droplets. More accurate experiments in immediate vicinity of the critical point are needed.

The size of the mesoscopic droplets increases with increasing hydrophobicity of the environment, upon approach to the conditions of macroscopic phase separation. Under these conditions, the phenomenon of mesoscale solubilization starts overlapping with the ouzo effect - the formation of micron-size droplets by nucleation followed by Ostwald ripening and macroscopic phase separation. What is the connection between the ouzo effect (which is a purely kinetic phenomenon, even relatively long-lived) and mesoscale solubilized droplets which are so long-lived that they practically can be viewed as an equilibrium phenomenon defined by thermodynamics? We still do not know a full answer to this question.

Importantly, the addition of small amounts of CHX and the existence of mesoscale droplets of solubilized CHX does not change the universal scaling power law for the correlation length within the accuracy of our experiment. However, the location of the critical point changes. This explains the fact that in spite of different values of the critical temperature reported by different investigators for the binary mixture 2,6-lutidine–water (as well as for many other near-critical binary solutions), there is consensus on the character of the critical behavior in agreement with scaling theory. However, at higher concentrations of the third component, when the experimental path is at constant composition, a phenomenon of renormalization of critical exponents is predicted by theory [61] and observed experimentally [62]. In particular, the exponent of the correlation length will be renormalized from the universal value $\nu = 0.63$ to a higher (but also universal) value, where $\alpha = 0.11$ is the critical exponent for the heat capacity [61,62]. The characteristic temperature, $(T_c - T_x)/T_c$, of crossover between these two asymptotic values can estimated (by the order of magnitude) as follows [62,63]:

$$\frac{T_c - T_x}{T_c} \cong \left[ x \left( \frac{dT_c}{dx} \right)^2 \right]^{1/\alpha}, \qquad (9)$$

where $x$ is the mole fraction of the third component. At the temperatures $T_c - T \ll T_c - T_x$ the exponent $\nu$ is renormalized to $\nu'$. Since $1/\alpha \cong 9$, the crossover temperature is extremely sensitive to the degree if dilution. For the ternary system 2,6-lutidine-H$_2$O-CHX $dT_c/dx \cong 10$. This gives for $x \cong 10^{-3}$ the crossover temperature $(T_c - T_x)/T_c \cong 10^{-9}$. However for $x \cong 5 \cdot 10^{-3}$ the renormalization would occur at about $2 \cdot 10^{-3}$, in the experimental range. Thus the systematic deviations of the data for 0.16 % (mass) CHX ($x < 10^{-3}$) are unlikely to be attributed to the beginning of renormalization of the power law.

## 5. Conclusions

In this work, we investigated the effects of the presence of small amounts of a hydrophobic compound (cyclohexane) on critical and phase behavior of binary and quasibinary aqueous solutions of hydrotropes (2,6-lutidine-H$_2$O, 2,6-lutidine-H$_2$O/D$_2$O and solutions of an effective alcohol, a mix of *tert*-butanol/2-butanol, in H$_2$O). The correlation length of critical fluctuations near the liquid-liquid separation and the size of mesoscopic droplets of solubilized cyclohexane were measured by dynamic light scattering and by small-angle contrast-matching neutron scattering. It was shown that the effect of the presence of small amounts of cyclohexane on the near-critical phase behavior is twofold: 1) the transition temperature shifts towards increasing the two-phase domain; 2) long-lived mesoscopic (order of a hundred nm) inhomogeneities emerge; they remain unchanged near the critical point of macroscopic phase separation and do not alter the universal nature of critical anomalies. A larger amount of a hydrotrope generates alternative liquid-liquid phase transition at lower temperatures. This transition is manifested by emulsification ("ouzo effect") following by Ostwald ripening and macroscopic phase separation. We were not able to observe effects of critical fluctuations on the mesoscopic droplets. More accurate experiments in immediate vicinity of the critical point are highly desirable.

**Acknowledgements and Funding Sources**

We appreciate fruitful collaboration with Dr. Deepa Subramanian. Research of E.V.J. and M.A.A. was supported by the IPST Light Scattering Center at the University of Maryland, College Park. In Moscow, the study was supported by the Russian Foundation for Basic Research, Grant No. 15-08-07727-a. C.E.B. acknowledges the support of the National Research Council. This work utilized facilities supported in part by the National Science Foundation under Agreement No. DMR-0944772. The identification of commercial products does not imply endorsement by the National Institute of Standards and Technology nor does it imply that these are the best for the purpose.